\documentclass[article,useAMS,usenatbib]{mn2e}
\usepackage{graphicx}
\usepackage{eufrak}
\usepackage{eucal}
\usepackage{txfonts}
\usepackage{xcolor}

\begin{document}

\title[On the (in)variance of the dust-to-metals ratio in galaxies]{On the (in)variance of the dust-to-metals ratio in galaxies}

\author[Mattsson]{Lars Mattsson$^{1,2}$, Annalisa De Cia$^{3}$, Anja C. Andersen$^1$ \& Tayyaba Zafar$^{4}$\\
$^1$Dark Cosmology Centre, Niels Bohr Institute, University of Copenhagen, Juliane Maries Vej 30, DK-2100, Copenhagen \O, Denmark\\
$^2$Nordic Institute for Theoretical Physics (Nordita), KTH Royal Institute of Technology and Stockholm University, Roslagstullsbacken 23, SE-106 91, Stockholm, Sweden\\
$^3$Department of Particle Physics and Astrophysics, Faculty of Physics, Weizmann Institute of Science, Rehovot 76100, Israel\\
$^4$European Southern Observatory, Karl-Schwarzschildstrasse 2, 85748 Garching bei M\"unchen, Germany
}

\date{}

\pagerange{\pageref{firstpage}--\pageref{lastpage}} \pubyear{2013}

\maketitle

\label{firstpage}

\begin{abstract}
Recent works have demonstrated a surprisingly small variation of the dust-to-metals ratio in different environments and a correlation between dust extinction and the density of stars. Naively, one would interpret these findings as strong evidence of cosmic dust being produced mainly by stars. But other observational evidence suggest there is a significant variation of the dust-to-metals ratio with metallicity. As we demonstrate in this paper, a simple star-dust scenario is problematic also in the sense that it requires that destruction of dust in the interstellar medium (e.g., due to passage of supernova shocks) must be highly inefficient. We suggest a model where stellar dust production is indeed efficient, but where interstellar dust growth is equally important and acts as a replenishment mechanism which can counteract the effects of dust destruction. This model appears to resolve the seemingly contradictive observations, given that the ratio of the effective (stellar) dust and metal yields is not universal and thus may change from one environment to another, depending on metallicity.
\end{abstract}

\begin{keywords}
Galaxies: evolution, spiral; Stars: AGB and post-AGB, supernovae: general; ISM: dust, extinction;
\end{keywords}

\section{Introduction}
The variation of the overall dust-to-metals ratios between galaxies of vastly different morphology, ages and metallicities appears surprisingly small in many cases, with a mean value close to the Galactic ratio ($\sim 0.5$). The relatively tight correlation between the dust-to-gas ratio and the metallicity (yielding an almost invariant dust-to-metals ratio) in the Local Group galaxies has been known for quite a while \citep[see][]{Viallefond82,Issa90,Whittet91}. Indirect evidence for a `universal' mean value is also provided by the almost linear relation between $B$-band optical depth and stellar surface density in spiral galaxies \citep{Grootes13}. But recent results based on gamma-ray burst (GRB) afterglows, quasar foreground damped Ly$\alpha$-absorbers \citep[DLAs;][]{Zafar13} and distant lens galaxies \citep[see, e.g.,][]{Dai09,Chen13} now seem to extend this correlation beyond the local Universe and down to metallicties just a few percent of the solar value. \citet{Zafar13} argue this can only be explained by either rapid dust enrichment by supernovae or very rapid interstellar grain growth by accretion of metals.

However, there can be significant variations {\it within} a galaxy \citep[see, e.g.,][]{Mattsson12,Mattsson12b}, although the existence of dust-to-metals gradients is somewhat difficult to establish observationally with reliable independent methods. If, on the other hand, the dust-to-metals ratio does not vary much at all, in any environment, one may assume dust grains as well as atomic metals are mainly produced by stars. Recent findings of large amounts of cold dust in supernova (SN) remnants \citep{Matsuura11,Gomez12} seem to support this hypothesis, although the exact numbers can be disputed \citep{Temim13,Mattsson13a,Mattsson13b}. In other words: the overall picture is not consistent.

A new study by \citet{DeCia13} seems to confirm the rising trend with metallicity of the dust-to-metals ratio in quasar DLAs found in previous studies \citep{Vladilo98, Vladilo04}. Furthermore, \citet[][see also Herrera-Camus et al. 2012]{Fisher13} derived a dust mass in the local starburst dwarf I Zw 18, as well as a high-redshift object of similar character, which clearly indicate a dust-to-metals ratio below the Galactic value. These results, together with the likely existence of dust-to-metals gradients along galaxy discs \citep{Mattsson12b}, suggest the variance (or invariance) of the dust-to-metals ratio may depend on the environment. In such case, there may exist an equilibrium mechanism that keeps the dust-to-metals ratio close to constant if certain conditions are fulfilled, while a metallicity dependence may occur as a result of deviations from those conditions in other environments.

Recently, \citet{Kuo13} have tried to alleviate the tension between the results from the GRB afterglows of \citet{Zafar13} and other data (for local dwarf galaxies) by fine-tuning the parameters of their standard galactic dust evolution model including grain growth \citep{Hirashita11, Kuo12}. What they suggest is a quite reasonable compromise, but a truly convincing explanation of the different trends (constant and rising dust-to-metals ratio) would require some modification of the dust-formation scenario. In particular, a model in which inherent properties of a galaxy more or less uniquely determines its dust-to-metals ratio would be desirable. Even if the models by \citet{Kuo13} are marginally consistent with the data they compare with, there is obviously still some tension between models and observations. The new results by \citet{DeCia13} only act as to emphasise this.  In the standard picture of production and destruction of cosmic dust one is faced with the following two problems: (1) in metal-poor environments dust is only supplied by stars as the interstellar density of metals is too low for efficient grain growth, but still being destroyed by SN shockwaves (albeit with a relatively low efficiency); (2) to compensate the destruction of dust grains, which eventually becomes efficient, with grain growth requires that one pushes the boundaries of the model, i.e., to obtain a sufficiently short grain-growth timescale, one is forced to accept a very large span of gas densities (several orders of magnitude) and a very low star-formation efficiency. These problems are discussed in more detail by \citet{Kuo13}.

In this paper we investigate a scenario for the evolution of the galactic dust component where destruction of grains due to sputtering in SN shockwaves is roughly balanced by grain growth by accretion of molecular gas. This idea has also been put forth in other studies to improve models of the build-up of dust in the local as well as distant (early) Universe \citep[see, e.g.,][]{Inoue11,Mattsson11,Valiante11}, but here we take it one step further and consider a model where there can be an {\it exact} balance. Given a constant ratio of the effective dust yield and the total metal yield for a generation of stars, such a scenario will lead to an invariant dust-to-metals ratio. We continue by discussing the possibility that young undeveloped (low metallicity) systems may have a different yield ratio due to different dust yields for individual stars (e.g., the expected metallicity dependence).

\section{Observational clues and constraints on the dust-to-metals ratio}
Recently, \citet{Grootes13} derived a correlation between the optical depth in the $B$-band $\tau_B$ and the stellar-mass surface density $\Sigma_\star$ in nearby spiral galaxies selected from the Galaxy and Mass Assembly (GAMA) survey, which were detected in the FIR/sub-mm in the Herschel-ATLAS field. They find a nearly linear relation,
\begin{equation}
\log(\tau_B) = (1.12\pm 0.11) \times \log\left({\Sigma_\star\over M_\odot\,{\rm kpc}^{-2}} \right) - 8.6\pm 0.8,
\end{equation}
where the errors reflect the $1\sigma$ scatter in the data. The regression is marginally consistent with an exactly linear correlation between $\tau_B$ and $\Sigma_\star$. If the dust density $\Sigma_{\rm d}$ is proportional to the stellar-mass density $\Sigma_{\rm stars}$, there should exist a linear correlation between the optical depth $\tau_{\lambda}$ and the corrected, de-projected surface density of a galaxy, much like the relation above, because of the connection with the dust abundance, i.e., $\tau_\lambda \sim \Sigma_{\rm d}$. \citet{Grootes13}, argue that their relation is evidence of efficient interstellar grain growth. This conclusion depends on whether stars, primarily massive stars, can make a significant contribution to the dust production and on the efficiency of dust destruction in the interstellar medium (ISM). In principle, the $\tau_B - \Sigma_\star$ connection only says that dust and stars 'go hand-in-hand' in local spiral galaxies, which seems to suggest significant stellar dust production and that destruction of dust must be balanced by grain growth in the ISM. However, as we will go on to show later, this is not necessarily the case.

A constant dust-to-metals ratio does not only seem to apply in the local Universe, however. \citet{Zafar13} combine extinction ($A_V$ values) and abundance data from GRB afterglows with similar data from QSO foreground absorbers and multiply-imaged galaxy-lensed QSOs, to determine the dust-to-metals ratios for a wide range of galaxy types and redshifts of $z = 0.1- 6.3$, and almost three orders in metal abundance. The mean dust-to-metals ratio for their sample is very close to the Galactic value and the $1\sigma$ deviation is no more than 0.3 dex, suggesting the dust-to-metals ratio may be fairly invariant throughout the observable Universe. {\it Chandra} X-ray observations of distant lens galaxies lend further support to this picture \citep{Chen13,Dai09}. Taken at face value, these results would imply a very rapid dust-formation scenario that is roughly the same in any environment.

The number of data points at low metallicity is relatively small in the work by \citet{Zafar13}. It is therefore not certain \textbf{that} the dust-to-metals ratio is nearly invariant also at low metallicities. A very recent study by \citet{DeCia13} has shown, using a different method, that there is likely a turn-down in the dust-to-metals ratio at low metallicity. This is also consistent with the constraint on the dust-to-metals ratio derived for the local starburst galaxy I Zw 18 \citep{Herrera-Camus12,Fisher13}. \citet{DeCia13} measure the degrees of depletion of gas-phase abundances in the ISM for various elements, particularly focusing on Fe and Zn, and infer the dust abundance from these depletions. The dust-depletion patterns are observed in UV/optical GRB afterglows and QSO spectra, associated with the ISM of the GRB host-galaxies and QSO-DLAs, and are derived assuming that the depletion is entirely due to dust condensation, regardless of its origin. In particular, the method used by \citet{DeCia13} relies on the assumption that the observed [Zn/Fe] traces the overall dust content in the ISM, and thus that (1) the intrinsic relative abundance of Zn and Fe is solar and (2) a non-negligible amount of iron is present in the bulk of the dust. This is not obviously the case due to uncertainties in the origins of Zn and Fe, but investigating the reliability of these assumptions - and thus the exact slope of the trend of the dust-to-metals ratio with metallicity - goes beyond the scope of this paper.

What is particularly interesting about the new results by \citet{DeCia13} is that the dust-to-metals ratio increases with increasing metallicity and, even more important, with increasing metal {\it density}.  The latter is a clear indication of grain growth being an important part of the build up of the dust mass. Further evidence from DLAs of a down-turn in the dust-to-metals ratio at low metallicity is seen in the works by, e.g., \citet{Vladilo98,Vladilo04}. A similar, although somewhat steeper, down-turn of the dust-to-\textit{gas} ratio was also recently found by \citet{Remy-Ruyer13} for low-metallicity galaxies in the local Universe.

\section{Dust processing in the ISM}

\subsection{Grain growth}
In a gaseous medium of a given temperature and density, the rate of accretion of a gas-phase species $i$ onto a spherical dust grain is given by the surface area of the grain ($4\pi a^2$ where $a$ is the grain radius) and the sticking coefficient (probability) $f_{\rm s}$ for that species \citep[see, e.g.][]{Dwek98}. The mass density of a species $i$ locked up in dust $\rho_{{\rm d}, i}$ then grows at a rate
\begin{equation}
{1\over \rho_{{\rm d},i}} {d\rho_{{\rm d},i}\over dt} = 3f_{\rm s} {\langle v\rangle \over a_{\rm eff}} {\rho_i-\rho_{{\rm d},i}\over \rho_{\rm gr}},
\end{equation}
where $\rho_i$ denotes mass density per unit volume of the growth species $i$, $\langle v_{\rm g}\rangle$ is the mean thermal speed of the gas particles, $a_{\rm eft}$ is the effective (average) grain size and $\rho_{\rm gr}$ is the material bulk density of the dust. Thus, the overall timescale of grain growth $\tau_{\rm grow}$ is, to first approximation, inversely proportional to the difference between total metallicity $Z$ and the dust-to-gas ratio $Z_{\rm d}$ and can therefore be approximated using \citep{Mattsson12}
\begin{equation}
\label{taugr}
\tau_{\rm grow} \propto {1\over Z\,\rho_{\rm H_2}}\left(1-{Z_{\rm d}\over Z}\right)^{-1},
\end{equation}
where $\rho_{\rm H_2} $ is the density of molecular hydrogen. Here, the grain size, sticking probability, thermal speed of the gas particles and their molecular composition have been regarded as more or less invariant quantities. 

For simplicity we will assume that the star formation rate is proportional to the molecular gas abundance. Thus,  $d\rho_{\rm s}/dt \propto \rho_{\rm H_2}$. We can then regard the timescale $\tau_{\rm grow}$ as essentially just a simple function of the metallicity, the gas abundance and the growth rate of the stellar component. Following \citet{Mattsson12} we adopt
\begin{equation}
\tau_{\rm grow} ^{-1}= {\epsilon Z \over\rho_{\rm g}} \left(1-{Z_{\rm d}\over Z}\right) {d\rho_{\rm s}\over dt},
\end{equation}
where the constant $\epsilon$ can be treated as a unit less free (but constrained) parameter of the model, representing the overall efficiency of grain growth. 

\subsection{Destruction by sputtering}
\label{destruction}
The dominant mechanism for dust destruction is by sputtering in the high-velocity interstellar shocks driven by SNe, which can be directly related to the energy of the SNe \citep{Nozawa06}.
Following \citet{McKee89,Dwek07} the dust destruction time-scale is
\begin{equation}
\tau_{\rm d} = {\rho_{\rm g}\over \langle m_{\rm ISM}\rangle\,R_{\rm SN}},
\end{equation}
where $\rho_{\rm g}$ is the gas mass density, $\langle m_{\rm ISM}\rangle$ is the effective gas mass cleared of dust by each SN event, and $R_{\rm SN}$ is the SN rate per unit volume. The latter may be approximated as
\begin{equation}
\label{snr}
R_{\rm SN}(t) \approx \dot{\rho}_{\rm sfr}(r,t)\int_{8M_\odot}^{100M_\odot} \phi(m)\,dm,
\end{equation}
where $\phi(m)$ is the stellar initial mass function (IMF) and $\dot{\rho}_{\rm sfr}$ is the star-formation rate per unit volume. For a non-evolving IMF the integral in equation (\ref{snr}) is a constant with respect to time, and is not expected to vary much spatially within a galaxy either. Hence, the time scale $\tau_{\rm d}$ may be expressed as
\begin{equation}
\label{taud}
\tau_{\rm d}^{-1} \approx  {\delta\over \rho_{\rm g}}{d\rho_{\rm s}\over dt},
\end{equation}
where $\delta$ will be referred to as the dust destruction parameter. This parameter is dimensionless, and as such it can be seen as a measure of the overall efficiency of dust destruction.

Small grains are more susceptible to destruction by sputtering in SN shock waves than large grains \citep{Slavin04}. This is due to the larger grains' tendency to decouple from the gas and thus being less exposed to ions. This fact suggests the above model is, partially, an inadequate description of the effects of destruction due to SN shocks. Grain-grain interaction may lead to shattering and thus creation of smaller grains \citep{Hirashita09,Asano13}, which are then more likely to be sputtered away. Hence, the timescale of dust destruction may not only be inversely proportional to the SN rate, but also the abundance of dust, since the rate of interactions (or collisions) is proportional to the number density $n_{\rm d}$.\footnote{ With the adaptations usually employed in chemical collision theory \citep{Atkins10}, the collision frequency is $R_{\rm coll} \equiv \sigma_{\rm coll} v_{\rm rel}\,n_{\rm d}$, where $\sigma_{\rm coll} = 2\pi \langle a^2\rangle$  is the effective cross-section for grain-grain collisions, $n_{\rm d}$ is the number density of dust grains and $v_{\rm rel}$ is the typical relative velocity of two colliding grains. Using $R_{\rm coll}$ we may define the {\it collision density} as ${1\over 2}R_{\rm coll}\,n_{\rm d}$. The factor $1/2$ has been introduced to avoid double-counting the collisions. Obviously, the collision density is proportional to $Z_{\rm d}^2$ since $n_{\rm d}\propto Z_{\rm d}$. The efficiency of dust destruction is roughly proportional to the shattering rate, since smaller fragments are more easily destroyed, and the shattering rate is to first order proportional to the collision rate, which sketchily motivates the modified model of dust destruction suggested above.} A reasonable modification to the dust-destruction timescale would then be to introduce a dependence on the dust-to-gas ratio $Z_{\rm d}$, i.e.,
\begin{equation}
\label{taud2}
\tau_{\rm d}^{-1} \approx  {\delta \over \rho_{\rm g}}{Z_{\rm d}\over Z_{\rm d,\,G}}{d\rho_{\rm s}\over dt},
\end{equation}
where $Z_{\rm d,\,G}$ is the present-day Galactic dust-to-gas ratio. 

The dust-destruction efficiency $\delta$ can be calibrated to the expected efficiency (timescale) for the Galaxy, which we take to be roughly 0.7 Gyr \citep{Jones96}. The effective Galactic gas-consumption rate is about $2\,M_\odot$~pc$^{-2}$~Gyr$^{-1}$, and the gas density is $\sim 8\,M_\odot$~pc$^{-2}$ \citep[see, e.g.,][and referenes therein]{Mattsson10}, which implies $\delta\approx 5$. \citet{Mattsson11} estimated $\delta\approx 10$ based on a \citet{Larson98} IMF and that stars of initial masses above $10\,M_\odot$ become SNe. We can thus assume $\delta \sim 5 - 10$ is a reasonable estimate of the expected range for $\delta$.

\section{Simple models of dust evolution}
In \citet{Mattsson12,Mattsson12b} we showed that dust growth would be the most important mechanism for changing the dust-to-metals ratio $\zeta$ in a galaxy throughout its course of evolution and/or create a dust-to-metals gradient along a galaxy disc. Since, in the present work, we want to also consider the situations where $\zeta$ is not changing much, we will focus on the two viable scenarios for dust production: (1) pure stellar dust production and inefficient dust destruction and (2) a scenario where dust destruction is balanced by dust growth in the ISM.  

To simplify our model we make the same assumptions as in \citet{Mattsson12} and \citet{Mattsson12b}, i.e., a galaxy evolves effectively as a `closed box' and the stellar dust/metals production can be described under the instantaneous recycling approximation. We also assume the effects of the inevitably chaining grain-size distribution are negligible on average, so that grain growth and destruction are functions of macroscopic properties only as described in the next subsection. Furthermore, we make the assumption that the fraction of condensible metals (metals that may end up in dust grains) $Z_{\rm c}$ is essentially the same as the total metallicity, i.e.,   $Z_{\rm c} \approx Z$. This assumption is quite reasonable as the observed depletion is surprisinglingy close to 100\% for many of the most abundant metals except oxygen and the noble gases \citep[see, e.g.,][]{Pinto13}. The equation for the evolution of the dust-to-metals ratio $\zeta = Z_{\rm d}/Z$ is then \citep{Mattsson12},
\begin{equation}
\label{dtm0}
Z{d\zeta\over dZ} =  {y_{\rm d}\over y_Z} + {\zeta Z\over y_Z}[G(Z)-D(Z)]- \zeta,
\end{equation}
where $G$ is the rate of increase of the dust mass due to grain growth relative to the rate of gas consumption due to star formation, $D$ is the corresponding function for dust destruction and $y_{\rm d}$, $y_Z$ is the effective stellar dust and metal yields, respectively. The dust yield $y_{\rm d}$ may have a significant dependence on the metallicity of the stellar population, which we will return to later. In terms of the timescales for grain growth and destruction above, $G$ and $D$ can be defined as
\begin{equation}
\label{GD}
G(Z) = \epsilon Z\,\left[1-{Z_{\rm d}(Z)\over Z}\right], \quad D = {\delta} \quad {\rm or}\quad D(Z) = \delta' Z_{\rm d}(Z),
\end{equation}
where $\delta/\delta' = Z_{\rm d,\,G}$.

\subsection{Pure stellar dust production}
\label{purestellar}
We first consider the case where we have only stellar dust production and no destruction of dust in the ISM ($\epsilon = \delta = 0$). For a `closed box', the dust-to-gas ratio $Z_{\rm d}$ is simply given by
\begin{equation}
Z_{\rm d} = y_{\rm d}\ln\left(1+{\Sigma_\star\over \Sigma_{\rm gas}} \right),
\end{equation}
Note that by replacing $y_{\rm d}$ with $y_Z$, we would obtain the corresponding relations for metallicity. Series expansion around $\Sigma_\star/\Sigma_{\rm gas} = 0$ yields 
 \begin{equation}
 \label{linear}
Z_{\rm d} = y_{\rm d}{\Sigma_\star\over \Sigma_{\rm gas}} + {y_{\rm d}\over 2}\left({\Sigma_\star\over \Sigma_{\rm gas}} \right)^2 - \dots ,
\end{equation}
from which we may conclude that $\Sigma_{\rm d}\approx y_{\rm d}\,\Sigma_\star$ for an unevolved galaxy where $\Sigma_\star/\Sigma_{\rm gas}\ll 1$. Thus, the dust masses in young starbursts, like I Zw18, should give us a measure of the stellar dust yield $y_{\rm d}$ (at least for low metallicities). This may also give a hint about the origin of the $\Sigma_{\rm d} \sim \Sigma_\star$ connection seen in the results by \citet{Grootes13}, i.e., that we should consider a model where a balance between growth and destruction leads to a similar $\Sigma_{\rm d}\sim \Sigma_\star$ relation for more evolved systems.

If we include interstellar dust destruction with a timescale given by eq. (\ref{taud}) ($D = \delta $, $G  = 0$) the closed-box solution to (\ref{dtm0}) can be written
in the form
\begin{equation}
\zeta = {Z_{\rm d}\over Z} = {y_{\rm d}\over y_Z}{1\over \delta}\left[1-\left(1+{\Sigma_\star\over \Sigma_{\rm gas}} \right)^{-\delta}\right]\ln\left(1+{\Sigma_\star\over \Sigma_{\rm gas}}\right)^{-1}.
\end{equation}
Analysis of this solution shows that $\Sigma_\star/\Sigma_{\rm gas}\gg 1$ requires $\zeta \ll1$ \citep{Mattsson11}. Using the timescale given by eq. (\ref{taud2}), which is based on the suggested grain-grain interactions ($D = \delta' Z_{\rm d} $, $G  = 0$), gives a solution of the form
\begin{equation}
\zeta = {Z_{\rm d}\over Z} = {1 \over y_Z}\sqrt{y_{\rm d}\over \delta'}\tanh \left[\sqrt{y_{\rm d} \delta'}  \ln\left(1+{\Sigma_\star\over \Sigma_{\rm gas}}\right)\right],
\end{equation}
which suggests the same asymptotic behaviour, i.e., $\Sigma_\star/\Sigma_{\rm gas}\gg 1$ requires $\zeta \ll1$. This tells us that only stellar dust production cannot work if there is interstellar dust destruction on any level after the dust has become part of the diffuse ISM. The dust-to-metals ratio $\zeta$ will decrease monotonously unless the effective stellar dust yield $y_{\rm d}$ increases in such a way that it compensates for the dust destruction. Otherwise, if we are to maintain a roughly constant $\zeta$, there cannot be any significant destruction of dust in the ISM.

\subsection{Growth/destruction equilibrium model}
\label{gdequil}
With $G$ as in Eq. (\ref{GD}) and $D=\delta$ (the `canonical' model of dust destruction) we have an equation for $\zeta$ which reads
\begin{equation}
\label{dtm}
Z{d\zeta\over dZ} =  {y_{\rm d}\over y_Z} + {\zeta Z\over y_Z}\left[\epsilon \left(1-\zeta \right)\,Z -\delta\right] - \zeta.
\end{equation}
The equilibrium case $d\zeta/dZ = 0$ would correspond to $\epsilon(1-\zeta)\,Z-\delta = 0$ and $\zeta = y_{\rm d}/y_Z$, which is equivalent to the criterion
\begin{equation}
\label{crit1}
{\delta\over\epsilon}= Z\,\left(1-{y_{\rm d}\over y_Z}\right).
\end{equation} 
This is a problem, however, since $y_{\rm d}$, $y_Z$, as well as $\delta$, $\epsilon$ are constants by definition, while $Z$ cannot be constant, except under very special conditions. It is therefore virtually impossible to keep $\zeta$ more or less constant over a wide range of metallicities.   

If we instead consider our second equation of dust evolution,
\begin{equation}
\label{dtm2}
Z{d\zeta\over dZ} =  {y_{\rm d}\over y_Z} + {\zeta Z^2\over y_Z}\left[\epsilon \left(1-\zeta \right) -\delta' \zeta\right] - \zeta,
\end{equation}
for the case where the dust-destruction timescale depends on the dust-to-gas ratio $Z_{\rm d}$, i.e., $D(Z) = \delta' Z_{\rm d}(Z)$, where $\delta' = \delta/Z_{\rm d,\,G}$, we obtain a more realistic equilibrium condition. More precisely, we have that $\epsilon(1-\zeta)-\delta' \zeta = 0$, which leads to
\begin{equation}
{\delta'\over\epsilon}= {y_Z \over y_{\rm d}}-1.
\end{equation} 
This criterion is more useful than Eq. (\ref{crit1}), since it does not involve any variable. If we adopt the Galactic dust-to-metals ratio, $\zeta_{\rm G}\approx 0.5$, we have $y_{\rm d}/y_Z\approx 0.5$ and thus $\epsilon \approx \delta'$. With $\delta \sim 5-10$ and $\delta/\delta' \approx 100$ (Galactic gas-to-dust ratio), we then have $\epsilon \sim 500-1000$, which suggests a relatively high efficiency of grain growth is required to only maintain balance between growth and destruction. A parameter range $\epsilon \sim 500-1000$ is consistent with the results by \citet{Mattsson12b}.

The special case $\epsilon = \delta'$ is worth some further consideration. Provided there is no dust if $Z=0$, it follows directly from Eq. (\ref{dtm2}) that $\zeta(0) = y_{\rm d}/y_Z$ regardless of whether $\epsilon = \delta'$ or not. In the opposite limit (large $Z$) the dust-to-metals ratio $\zeta$ will approach its asymptotic value and thus be constant. Hence, Eq. (\ref{dtm2}) reduces to
\begin{equation}
0 = {\epsilon \zeta\over y_Z} (1-2\zeta),
\end{equation}
which corresponds to $\zeta\to 1/2$ (the asymptotic value). Thus, if $\epsilon$ and $\delta'$ are similar, regardless of the actual value, we would have $\zeta \sim 0.5$. This result is particularly interesting since the dust-to-metals ratio in essentially all Local Group galaxies are close to $\zeta \approx 0.5$ \citep{Inoue03,Draine07}. With the model suggested above, this ratio would be a universal ratio which all galaxies will evolve toward, while the dust-to-metals ratio at early times may be quite different. A similar idea is discussed in \citet{Inoue11}.

The general solution to Eq. (\ref{dtm2}) for the initial condition $Z_{\rm d}(0) = Z(0) = 0$ and $\epsilon > 0$ is 
\begin{equation}
\label{solution1}
\zeta = {y_{\rm d}\over y_Z} {
M\left[1+{1\over 2}{y_{\rm d}\over y_Z}\left(1+ {\delta'\over \epsilon} \right), {3\over 2}; {1\over 2}{\epsilon Z^2\over y_Z}\right]
\over 
M\left[{1\over 2}{y_{\rm d}\over y_Z}\left(1+ {\delta'\over \epsilon} \right), {1\over 2}; {1\over 2}{\epsilon Z^2 \over y_Z}\right]}, 
\end{equation}
where $M(a,b;z)$ is the Kummer-Tricomi function of the first kind, which is identical to the confluent hypergeometric function $_1F_1(a,b;z)$. \citep[see][for proof that Eqns. \ref{dtm} and \ref{dtm2} can be transformed into Kummer's equation]{Mattsson12}. The growth/destruction equilibrium case, ${\delta' /\epsilon}= {y_Z / y_{\rm d}}-1$, corresponds to $a=b$, where we note that $M(a,a;z) = e^z$. Consequently, $\zeta = y_{\rm d}/y_Z$, as discussed above. In reality, one would expect deviations from an exactly constant dust-to-metals ratio to occur as a consequence of local variations of the yield ratio $y_{\rm d}/y_Z$ together with $\delta'$ and $\epsilon$. The latter two parameters are clearly different for different dust compositions and may also have implicit dependences on the gas density and, perhaps most importantly, on the grain-size distribution, which can only be `universal on average'.

For the case $\epsilon = \delta' = 0$ (neither dust growth, nor destruction) we have the trivial solution $\zeta = y_{\rm d}/y_Z$, which is of course identical to the equilibrium case above.

\subsection{Metallicity-dependent stellar dust production}
\label{z-dependence}
The effective stellar dust yield $y_{\rm d}$ has so far been treated as a constant. To first order, this is an acceptable approximation, but as we are here interested in dust production at very low metallicity it is necessary to consider a scenario in which  $y_{\rm d}$ is a function of the metallicity $Z$. There are two reasons for this. First, some key-elements for dust production (such as silicon) may be less abundant in low-metallicity stars. This is obviously the case for the massive, short-lived, AGB stars which are producing mainly silicates, but has no (or very little) silicon production of their own. Second, dust condensation is strongly dependent on the absolute abundance/density of the relevant metals. That is, there may exist a critical metallicity below which dust condensation become inefficient due to low partial pressures for many metals, leading to less nucleation and slow accretion. It is already well established that such a critical metallicity exists for grain growth in the ISM \citep[see, e.g.,][]{Asano13a} This can be the case also in massive stars which, despite that they produce significant amounts of metals, may have too low partial pressures of certain key-elements to have efficient nucleation. 

\begin{figure}
\resizebox{\hsize}{!}{
\includegraphics{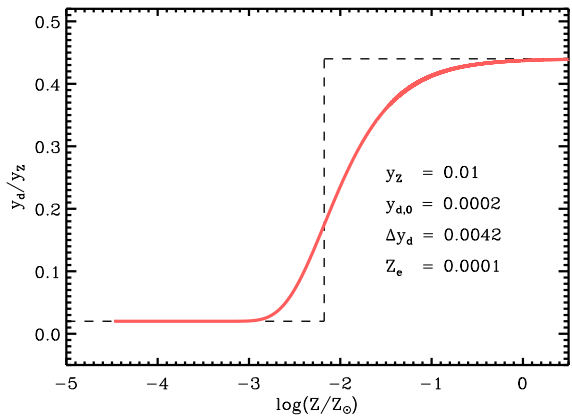}}
\caption{\label{dustyield} Effective stellar dust yield as a function of metallicity. The solid red line shows the smooth `jump' from low to high degree of dust condensation according to Eq. \ref{smooth}.}
\end{figure}

A very simple scenario would be the one where $y_{\rm d}$ is simply proportional to the metallicity $Z$. Assuming interstellar dust processing has no effect on the dust mass fraction  of the ISM ($G=D=0$ or $G=D\neq 0$) and $y_{\rm d}(Z) = y_{\rm d,0} + k\,Z$, where $y_{\rm d,0}$, $k$ are constants, we have
\begin{equation}
Z{d\zeta\over dZ} =  {y_{\rm d,0} + k\,Z\over y_Z} - \zeta,
\end {equation}
which has the simple solution [with initial condition $\zeta(0)=0$] 
\begin{equation}
\zeta(Z) = {1\over 2 }{y_{\rm d,0} + y_{\rm d}(Z)\over y_Z}. 
\end{equation}
This model produces a rising trend as seen in several observations, but is otherwise not very realistic. First, there is no `roof' in the solution above. $\zeta$ can continue to grow even beyond the absolute upper limit $\zeta = 1$. Second, it is expected that there is critical/threshold metallicity for efficient dust formation rather than a linear rise as above. Thus, a more realistic scenario is that in which stellar dust production becomes efficient at a certain metallicity, i.e., there is a smooth `jump' in $y_{\rm d}$ at some metallicity $Z_{\rm e}$. The transition from inefficient to efficient dust condensation is likely smooth, so it would be reasonable to adopt something of the form (see Fig. \ref{dustyield})
\begin{equation}
\label{smooth}
y_{\rm d}(Z) = y_{\rm d, 0} + \Delta y_{\rm d}\exp\left(-{Z_{\rm e}\over Z}\right),
\end {equation}
where $y_{\rm d, 0}$ is the minimum dust yield for inefficient dust condensation and $y_{\rm d, max} = y_{\rm d, 0} + \Delta y_{\rm d}$ is the maximum dust yield obtained at high efficiency. Thus, we obtain the solution [with initial condition $\zeta(Z_0)=\zeta_0$]
\begin{equation}
\zeta(Z) = {y_{\rm d}(Z) \over y_Z} + {\Delta y_{\rm d}\over y_Z}{Z_{\rm e}\over Z}\left[ {\rm E}_1\left({Z_{\rm e}\over Z_0} \right) - {\rm E}_1\left({Z_{\rm e}\over Z} \right) - {Z_0\over Z_{\rm e}}\exp\left(-{Z_{\rm e}\over Z_0}\right)\right]
\end{equation}
where we have defined the so-called exponential integral ${\rm E}_n$ as
\begin{equation}
{\rm E}_n(x) \equiv \int_1^\infty \frac{e^{-xt}}{t^n}\, dt.
\end{equation}
If the initial metallicity $Z_0$ is very small, or, more precisely, if $Z_0\ll Z_{\rm e}$, we can simplify the above expression into 
\begin{equation}
\label{z-dep-analytic}
\zeta(Z) = {y_{\rm d}(Z) \over y_Z} - {\Delta y_{\rm d}\over y_Z} {Z_{\rm e}\over Z} {\rm E}_1\left({Z_{\rm e}\over Z} \right).
\end{equation}

\section{Results and discussion}
Below we consider the dust-to-metals trends derived from optical depth, extinction magnitude and depletion levels of certain metals and compare them with the simplistic models described in the previous section. Furthermore, we present simple Monte Carlo simulations to demonstrate how such simple scenario would appear when allowing the model parameters to vary within a certain parameter space. 

\subsection{B-band optical depth and dust abundance: do stars dominate cosmic dust production?}
Due to the approximate proportionality between $A_V$ and the dust-to-gas ratio one would  expect the $B$-band optical depth $\tau_B$ to be a simple function of dust density. More precisely, $\tau_B \sim \Sigma_{\rm d}$. Given the result by \citet{Grootes13}, the dust mass density $\Sigma_{\rm d}$ is then simply proportional to the stellar mass density $\Sigma_{\star}$. Theoretically, this proportionality is expected for all unevovled (gas-rich) galaxies (see Section \ref{purestellar}). But for it to hold also for more evolved galaxies, a balance between growth and destruction of dust in the ISM is necessary.

The trend obtained by \citet{Grootes13} is fundamentally an empirical result, and consistent with a simple model where dust is produced by stars. Nevertheless, the nearly linear $\tau_B - \Sigma_{\star}$ is not answering the question whether dust is formed mainly in stars or grown in the ISM (growth and destruction can conspire to produce a $\Sigma_{\rm d}\sim \Sigma_\star$ relation), but seems to favour models with significant stellar dust production.

The linear relation discussed above may suggest the dust-to-metals ratio is not showing large variations since the metal content of a galaxy is typically correlated with the stellar mass \citep[see, e.g.,][]{Lara-Lopez13,Pilyugin13}, but there is still plenty of room for scatter and the relation is derived for local spiral galaxies which may be in similar evolutionary states where the dust-to-metals ratio has reached an `equilibrium plateau'. A more diverse sample of objects would therefore provide a more useful statistical constraint.

\subsection{Invariant dust-to-metals ratio?}
\begin{figure}
  \resizebox{\hsize}{!}{
   \includegraphics{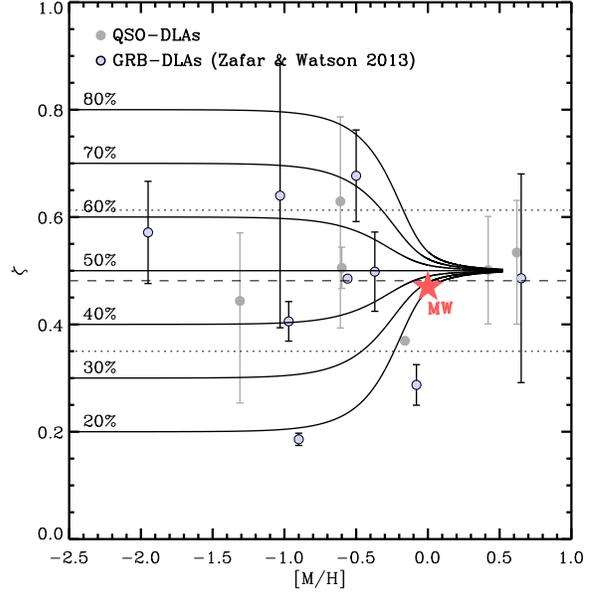}}
  \caption{\label{AV} Dust-to-metals ratio as a function of metallicity for a subset of the GRB and QSO-DLA sample and three QSO-DLAs used by \citet{Zafar13}. The mean ratio (dashed line) is essentially identical to the Galactic dust-to-metals ratio. The over-plotted 
  full-drawn lines show models with $\epsilon = \delta' = 750$ and various (constant) $y_{\rm d}/y_Z$ ratios ranging from $20-80$\%.}
  \end{figure}

We have transformed the dust-to-metals ratios in \citet{Zafar13} from observational units to unit less ratios\footnote{Defining the dust-to-metals ratio in observational units as $k/Z\equiv \log(N_{\textsc{Hi}}) + [{\rm X/H}] + \log(A_V)$, where $N_{\textsc{Hi}}$ is the column density of neutral hydrogen and [X/H] is the abundance of X relative to the corresponding solar value, we adopted the Galactic value $(k/Z)_G = 21.3$ \citep{Zafar13}. The unit less dust-to-metals ratio is obtained as $\zeta = \zeta_G 10^{[k/Z-(k/Z)_G]}$, where $\zeta_G \approx 0.5$. Here, we adopt $\zeta_G = 0.47$ to maintain consistency between the data sets. But the exact value is not very important as long as the adopted value is the same for all data sets considered.} (as in the models discussed in previous sections) for those objects where all relevant quantities have been measured with sufficient accuracy (see Fig. \ref{AV}). The relatively small variation of the dust-to-metals ratio ($\zeta = 0.47\pm 0.13$) seen over such a wide range of redshifts and metallicities (and likely also galaxy types) in the work by \citet{Zafar13} is a relatively strong constraint on the dust formation scenario, provided it can be trusted despite the rather small number of reliable measurements. A dust-to-metals ratio $\zeta = 0.47$ would correspond to $y_{\rm d}/y_Z = 0.47$ in an `equilibrium model' where $\delta' = \epsilon\,(y_Z/y_{\rm d}-1)$ (see Section \ref{gdequil}).  As shown by the different models (analytic solutions to Eq. \ref{dtm}) over-plotted in Fig. \ref{AV}, variation of the yield ratio $y_{\rm d}/y_Z$ leads to a wide range of dust-to-metals ratios at low metallicity, but converges to the asymptotic value, which is $\zeta = 0.5$ for the special case $\delta' = \epsilon$. The $1\sigma$ scatter in the \citet{Zafar13} data suggest $\zeta$ can vary at most about 30\%, but it should be noted the observed values cover a range $\zeta = 0.18 - 0.67$, which indicates significant variations of $\zeta$ cannot be ruled out due to small-number statistics. 

The growth/destruction-equilibrium model suggested in Section \ref{gdequil} is attractive as it would explain the existence of a characteristic, essentially universal, dust-to-metals ratio $\zeta$, as suggested by \citet{Zafar13}. Deviations from this `universal' value could then be attributed to variations of the yield ratio $y_{\rm d}/y_Z$. As we have mentioned in Section \ref{z-dependence}, there could exist a critical metallicity (or, more precisely, number density of certain key elements) in stars below which dust condensation is inefficient. However, it could also be that  $y_{\rm d}/y_Z$ is a `universal constant'  and that the limited variance in $\zeta$ could be explained by the fact that the growth and destruction parameters, $\epsilon$ and $\delta'$, respectively, can vary between different environments. Realistically, none of these parameters ($y_{\rm d}$, $y_Z$, $\epsilon$, $\delta'$) should be viewed as `universal constants', of course. We will return to this aspect of the variance in $\zeta$ in Section \ref{montecarlo}.

  \begin{figure}
  \resizebox{\hsize}{!}{
   \includegraphics{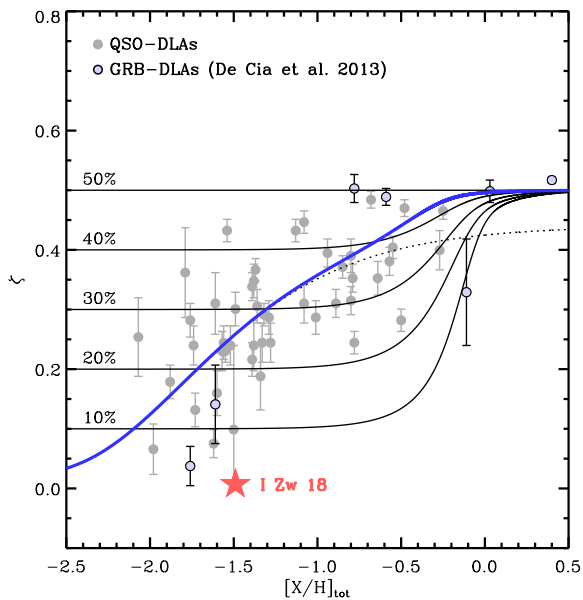}}
  \caption{\label{dep} Dust-to-metals ratio as a function of metallicity for a subset (objects with silicon-based metallicities were excluded) the GRB and QSO-DLAs considered by \citet{DeCia13}. The overall 
  trend is consistent with the dust-to-metals ratio derived for I Zw 18 by \citet{Fisher13}. The over-plotted solid black lines show models with $\epsilon = \delta'$ and various $y_{\rm d}/y_Z$ ratios ranging from 
  $10-50$\%. The blue line (grey in printed version) shows the best numerical solution with metallicity-dependent stellar dust yield, including grain processing in the ISM (see Sect. \ref{z-dependence}), and the 
  dotted line shows the corresponding (analytical) solution for the case of only metallicity-dependent stellar dust production (Eq. \ref{z-dep-analytic}). Note how the transition from stellar dust production to interstellar dust growth appears to
  happen at roughly 1/10 of solar metallicity. }
  \end{figure}

\subsection{Increasing dust-to-metals ratio?}
At first glance, an invariant dust-to-metals ratio in one context \citep[e.g.,][]{Grootes13,Zafar13} seem to be inconsistent with a clearly rising trend with metallicity in another \citep[e.g.,][]{DeCia13}. But as we have already discussed, the growth/destruction-equilibrium model with $\epsilon = \delta'$ has an asymptotic dust-to-metals ratio $\zeta_{\rm A}$ which is eventually reached regardless of what $\zeta$ is at early times. But if $\zeta$ shows a clear trend with metallicity, as in the results by \citet{DeCia13}, there cannot just be random variations of the yield ratio $y_{\rm d}/y_Z$. \citet{DeCia13}, as well as, e.g., \citet{Vladilo98}, find a $\zeta$ increasing with metallicity, which is what one would expect in a scenario where the bulk of cosmic dust is grown in the ISM rather than produced directly by stars. 

However, according to our simplistic model with a constant $y_{\rm d}$, the expected trend without growth/destruction-equilibrium is a {\it steep} rise in $\zeta$ at some critical metallicity \citep[see also][]{Mattsson12}, which is not in agreement with the observed trend (see Fig. \ref{dep}, models with $y_{\rm d}/y_Z < 0.5$). The observed slower rise of the dust-to-metals ratio can thus be a result of a changing yield ratio. If $y_{\rm d}$ increases at a certain metallicity, as described in Section \ref{z-dependence}, the observed trend could easily be explained. The analytic solutions for different $y_{\rm d}/y_Z $ (and $\epsilon = \delta'$) over-plotted in Fig. \ref{dep} show that if the yield ratio changes from a few percent at very low metallicity to $\sim 0.5$ at moderately low metallicity ($Z\sim 0.1\,Z_\odot$), the correct rising trend would be obtained. Ultimately, this shows that we need to modify our model - a constant yield ratio $y_{\rm d}/y_Z$ fails to reproduce the trend.

The blue line in Fig. \ref{dep} is a numerical solution (forth-order Runge-Kutta) using Eq. (\ref{smooth}) with the parameter values plotted in Fig. \ref{dustyield} to describe $y_{\rm d}(Z)$, which demonstrates exactly this point. At the same time, there is always a $y_{\rm d}/y_Z $ that will lead to a constant dust-to-metals ratio $\zeta$ for any given $\epsilon/\delta'$. We suggest this could be a good compromise in order to obtain a model that can explain why $\zeta$ in some cases show very little variation and in other cases a trend with metallicity. The case where interstellar dust processing has no effect on the dust mass fraction  of the ISM ($G=D=0$ or $G=D\neq 0$) is indicated by the dotted black line in Fig. \ref{dep} (corresponding to Eq. \ref{z-dep-analytic}). The effect of interstellar grain growth is the difference between the solid blue and dotted black lines, where the critical metallicity (the point where the lines diverge) occurs at $Z/Z_\odot \approx 0.1$. 

The most likely cause for a changing effective dust yield $y_{\rm d}$ is the existence of a critical metallicity below which dust formation is significantly less efficient compared to the efficiency at higher metallicities. As we have already mentioned, a lower number density of key-elements for dust condensation may be important in stars that do not produce much of these key-elements themselves. But for most massive stars that undergo a core-collapse supernova explosion the amount of metals produced is significant even at $Z=0$ \citep[see, e.g.,][and references therein]{Nomoto13}. However, gas opacities and cooling rates may be lower at very low metallicities, which in turn may affect the heating and cooling of existing dust grains. If the average grain temperature is high enough for sublimation to occur, the net efficiency of condensation may be low. Thus, it is not clear that very metal-poor stars can be efficient dust producers even if raw material for dust formation is present.   

Asymptotic giant branch (AGB) stars are probably not very important dust producers at low metallicity according to recent work in which a steep dependence on metallicty is found \citep{Ventura12}. In addition, at really low metallicity of the interstellar gas, i.e., at very early times, low- and intermediate-mass stars have not had enough time to evolve into AGB stars either. For example, metal-poor halo stars in the Galaxy appear to have been formed from gas that is mainly enriched by massive stars (supernovae with progenitor masses typically in the range $10-20\,M_\odot$), although variations in the abundance patterns sometimes occur \citep{Gilmore98}. Moreover, the destruction of dust in SNe is likely more efficient the more massive the progenitor star is (and the degree of dust condensation is likely lower), which means that a bias towards more massive stars at low metallicity may also lead to less stellar dust per unit stellar mass.  Numerical models of SN dust production do indeed confirm that the most massive stars have less surviving dust in their ejecta \citep{Bianchi07}. To summarise the above: {\it oxygen-rich AGB stars (the more massive and short-lived ones) cannot produce very much dust at low metallicity since they do not produce the refractory elements needed for dust production, and the effective dust yield of massive stars is probably strongly metallicity dependent too}. Thus, a $y_{\rm d}/y_Z $ increasing with metallicity seems reasonable.

The reason why the GRB and QSO-DLAs studied by \citet{Zafar13}, as well as local galaxies, show so little variation in their dust-to-metal ratios (despite a wide range of metallicities) is still not obvious. But provided the effective dust yield $y_{\rm d}$ depends on the metallicity, this invariant ratio as well as the rising trend found in quasar DLAs by measuring depletions \citep{Vladilo98,Vladilo04,DeCia13} could be `two sides of the same coin'. Statistical variations in the overall efficiencies of grain growth and destruction in the ISM, combined with some uncertainty in which metallicity $Z_{\rm e}$ stellar dust production starts to become efficient, will allow for enough scatter in the dust-to-metals ratio as a function of metallicity to have one fundamental model which is consistent with both the flat and the rising trend. This will be explored in the next section. As an alternative  hypothesis, one may consider the possibility that the $A_V$-based dust abundance estimates in \citet{Zafar13} are biased towards environments which have, relatively speaking, significant foreground contamination from intervening systems and therefore appear to have higher dust-to-metals ratios at low metallicity. This possibility should of course be investigated, but goes beyond the scope of this paper.

    \begin{figure*}
   \resizebox{\hsize}{!}{
   \includegraphics{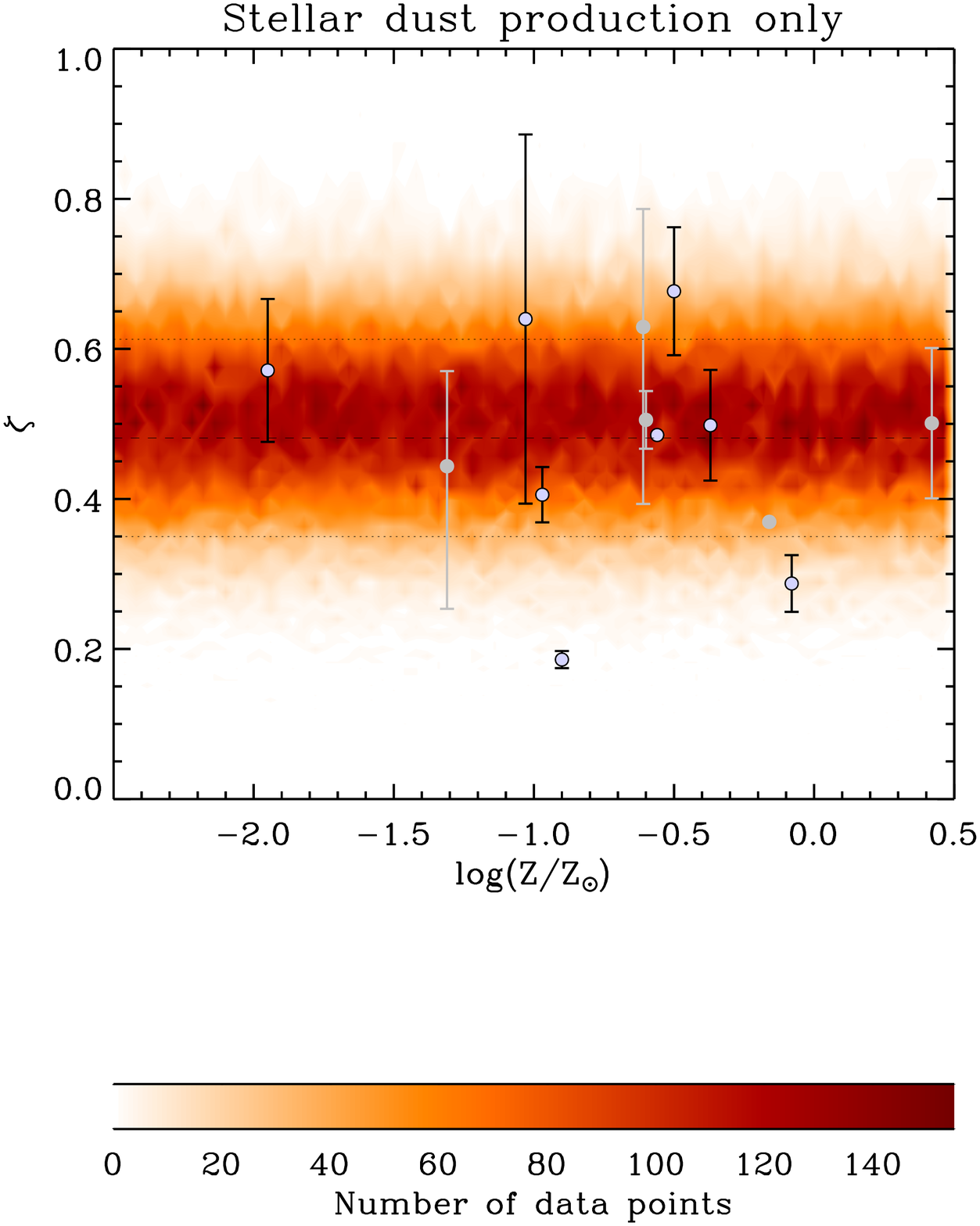}
   \includegraphics{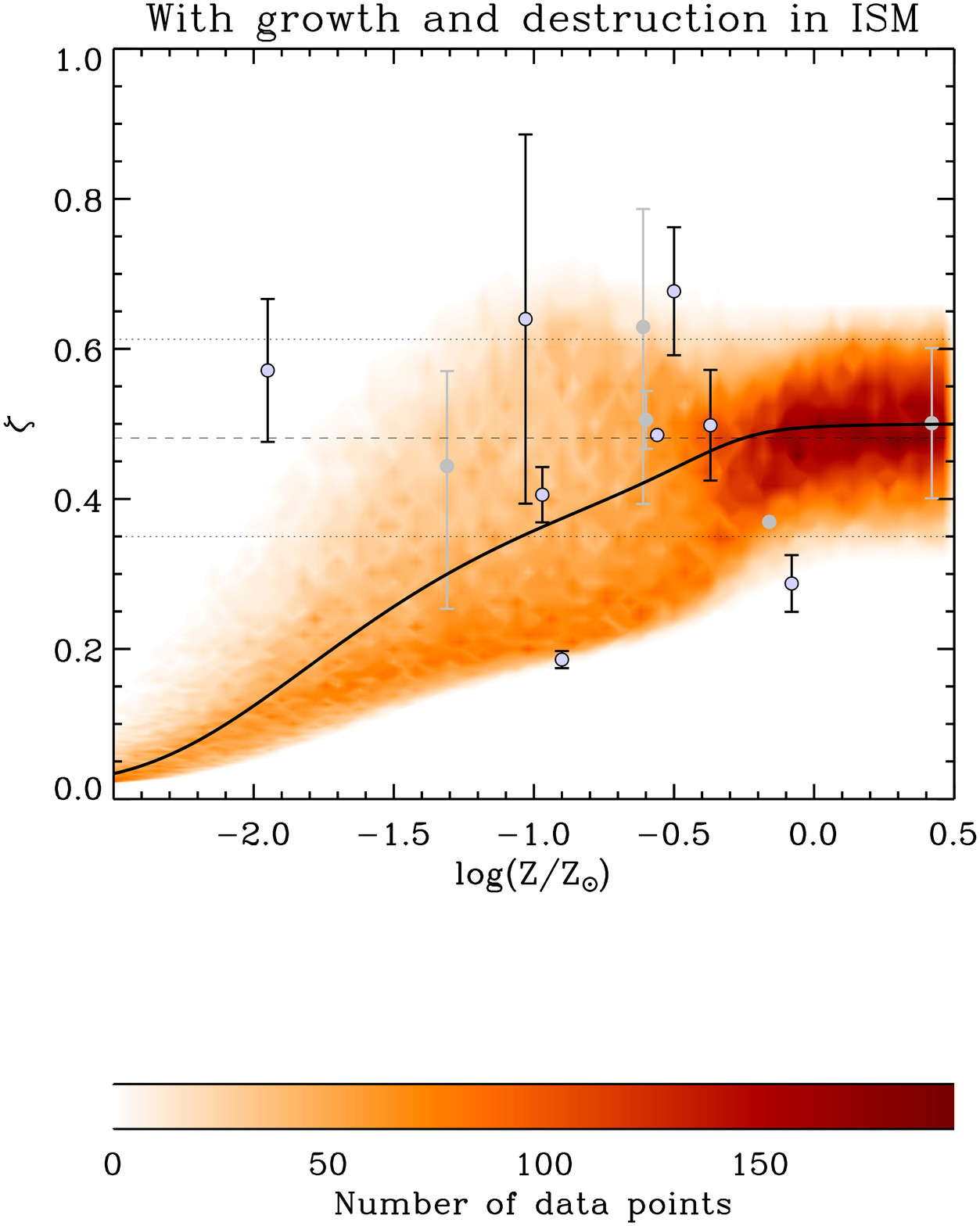}}
  \caption{\label{dtm_mc} Left panel: Monte Carlo simulation with stellar dust production and no interstellar growth and/or destruction. Model parameters (random variables) according to Table \ref{parameters}. 
  Right panel: same as the left panel but with interstellar growth and destruction included as well. The solid black line shows the same numerical solution as in Fig. \ref{dep}. The over plotted observational data is taken from \citet{Zafar13}.}
  \end{figure*}

  \begin{table}
  \caption{\label{parameters} Random variables/parameters used for the Monte Carlo models.}
  \center
  \begin{tabular}{llccc}
 Model & Variable& Mean &Range/std. dev. & Distribution\\
    \hline
  A: & $\log(Z/Z_\odot)$ & $-$ & $-2.5 \dots 0.5$ & Uniform \\  
  &$y_{\rm d}/y_Z$ & $0.5$ & $\pm 0.1$ & Normal \\
  &$\epsilon$ & $0$ & $-$ & $-$ \\
  &$\delta'$ & $0$ & $-$ & $-$ \\[1mm]
 B: & $\log(Z/Z_\odot)$ & $-$ & $-2.5 \dots 0.5$ & Uniform \\  
  &$Z_{\rm e}$ & $1.0\cdot 10^{-4}$ & $(0.75\dots 1.5)\cdot 10^{-4}$ & Uniform \\
  &$y_Z$ & $0.01$ & $0.005 \dots 0.015$ & Uniform \\
  &$\epsilon$ & $750$ & $500 \dots 1000$ & Uniform \\
  &$\delta'$ & $750$ & $500 \dots 1000$ & Uniform \\
  \hline
  \end{tabular}
  \end{table}

\subsection{Monte Carlo simulation of the dust-to-metals ratio as a function of metallicity}
\label{montecarlo}
We expect variations in not only the effective dust yield $y_{\rm d}$, but also in the timescales of grain growth and destruction ($\epsilon$ and $\delta$, in practice). To quantify the effects of such variations, to some extent, we have performed a couple of Monte Carlo simulations where we vary the parameters $\delta'$ and $\epsilon$ within reasonable ranges as well as setting them to zero (see Table \ref{parameters}). The yield ratio $y_{\rm d}/y_Z$ is not completely arbitrary either. On the one hand, the fraction of metals being injected into the ISM in the form of dust grains cannot be 100\%, since the degree of dust condensation must be limited by the physical conditions and the abundances of certain key elements (e.g., carbon or silicon) in the dust chemistry. On the other hand, this fraction cannot be too small either, since it is an observational fact that low- and intermediate-mass stars as well as massive stars in the local Universe produce significant amounts of dust. The fraction of dust that actually survive and eventually enrich the ISM is not known, but with the observed trend shown in Fig. \ref{dep} as reference we have calibrated the range of the effective yield ratio $y_{\rm d}/y_Z$ to approximately $0.02 - 0.44$. Thus, two of  the parameters of Eq. (\ref{smooth}) are fixed: $y_{\rm d,0} = 2.0\cdot 10^{-4}$ and $\Delta y_{\rm d} = 0.042$, while $Z_{\rm e}$ remains as a random variable of the Monte Carlo simulation together with $\delta'$ and $\epsilon$.

In Fig. \ref{dtm_mc} we have plotted the resultant probability density functions (PDF) of our simulation results.  To begin with, we performed a Monte Carlo simulation of the case of stellar dust production only, with the yield ratio $y_{\rm d}/y_Z$ ($y_{\rm d}$ not metallicity dependent) and the metallicity $Z$ as the only random variables. For this simulation we assumed that  $y_{\rm d}/y_Z$ follows a normal distribution with standard deviation $0.1$, centred at  $y_{\rm d}/y_Z = 0.5$ (model A in Table \ref{parameters}). The resultant PDF is consistent with data from \citet{Zafar13}, as can be seen in the left panel of Fig. \ref{dtm_mc}. After establishing this `bench mark', we then considered the case of a metallicity dependent dust yield according to Eq. (\ref{smooth}) with the parameter values given above and $Z_{\rm e} = 0.75-1.5\cdot10^{-4}$. The $\epsilon$ and $\delta'$ ranges are difficult to define, but as we argued in Section \ref{destruction}, $\delta \sim 5 - 10$ ($\delta'\sim 500-1000$) is a reasonable estimate of the expected range for $\delta$. Under the assumption $\epsilon \approx \delta'$, we may then assume $\epsilon \sim 500 - 1000$ (see model B in Table \ref{parameters}). All random variables were in this case assumed to follow uniform distributions.

As we showed in Section \ref{gdequil}, the dust-to-metals ratio converges to $\zeta = 0.5$ if $\epsilon = \delta'$, regardless of the value of $y_{\rm d}/y_Z$ or whether $y_{\rm d}$ is metallicity dependent or not. Clearly, this is the reason why the scatter in $\zeta$ becomes smaller at high metallicity when interstellar grain growth and destruction is included, compared to the case where $\epsilon = \delta' = 0$ in which the scatter is the same regardless of metallicity (cf. left and right panels in Fig. \ref{dtm_mc}). This inherent property of the model suggests one could, in principle, use the amount of scatter at approximately solar metallicity to constrain the width of the range of likely $\epsilon$ and $\delta'$ values. The observational data suggest a relatively small scatter (see Figs. \ref{AV} and \ref{dep}), albeit with large error bars on some data points. The parameter ranges that we have used in our simple Monte Carlo simulation appears to give a result that is consistent with the spread and uncertainty of the data at solar-like metallicities. Of course, one cannot draw very firm conclusions from a simplistic simulation like the present, but it seems that models which include interstellar grain growth and destruction is favoured by the fact that there appears to be significantly more scatter among the data points at low ($\sim 1/10$ of solar) metallicity than near solar metallicity. We therefore think our growth/destruction equilibrium model is plausible and may provide guidance towards a more consistent picture of the of the origin and evolution of cosmic dust.

\section{Conclusions}
Several observational studies suggest a surprisingly small variation of the dust-to-metals ratio in vastly different environments. It is worth stressing that the `trivial solution' to the problem, i.e., adopting a (constant) yield ratio of $y_{\rm d}/y_Z \sim 0.5$, works for any model where there is a replenishment mechanism to counteract possible dust destruction \citep[such as the model used by][for example]{Kuo13}. But other observational evidence also suggest there is a significant variation of the dust-to-metals ratio between different environments, and an invariant dust-to-metals ratio is problematic also in the sense that it requires fine-tuning and is pushing the limits of the `standard models' of dust evolution in galaxies to explain all data \citep{Kuo13}. 

We find that a reasonable way to resolve this apparent contradiction, and avoiding fine-tuning and extreme model parameters, is to assume that stellar dust production can be efficient, but  that interstellar dust growth is equally important and act as a replenishment mechanism which can almost exactly counteract the dust destruction in the ISM. In this scenario, the ratio of the effective (stellar) dust and metal yields is not likely a universal constant and may change due to some metallicity-dependence of the stellar dust yield. We propose the existence of a critical stellar metallicity above which nucleation and condensation of dust in stars can be efficient. 

We conclude that destruction and growth of grains in the ISM likely strives towards an equilibrium state, which mimics the general behaviour of the case of pure stellar dust production (and no destruction of grains). This explains the relatively small variation of the dust-to-metals ratio seen in several observational studies of local galaxies, but allows also for a significantly lower ratio at low metallicity if the effective stellar dust yield can vary with metallicity. 

The suggested scenario has important implications for the rapid build-up of large dust masses at high redshifts. Instead of requiring an extreme efficiency of dust formation in massive stars (SNe) as suggested by, e.g., \citet{Dwek07}, the large dust masses seen in the quasar-host galaxy SDSS J1148+5251 (and other objects at high redshifts), follows naturally from the rapid production of metals that is expected in a massive starburst. Just as \citet{Valiante11} we are led to conclude that, though massive stars must produce significant amounts of dust, dust masses of the order $10^8-10^9\,M_\odot$ (as in SDSS J1148+5251) are not likely a result of stellar dust sources only (as a consequence of interstellar dust destruction) and the resultant dust component must therefore be dominated by grain growth in molecular clouds.  
 
\section*{Acknowledgments}
The authors thank the anonymous reviewer for his/her constructive criticism which helped improve this paper.
Nordita is funded by the Nordic Council of Ministers, the Swedish Research Council, and the two host universities, the Royal Institute of Technology (KTH) and Stockholm University. The Dark Cosmology Centre is funded by the Danish National Research Foundation. ADC acknowledges support by the Weizmann Institute of Science Dean of Physics Fellowship and the Koshland Center for Basic Research.

\end{document}